# Linkage-length dependent structuring behavior of bent-core molecules in helical nanostructures

Hanim Kim,[a] Anna Zep,[b] Seong Ho Ryu,[a] Hyungju Ahn,[c] Tae Joo Shin,[d]

Sang Bok Lee,[e] Damian Pociecha,[b] Ewa Gorecka*[b] and Dong Ki Yoon*[a]

[a] *Graduate School of Nanoscience and Technology, KAIST, Daejeon 305-701, Republic of Korea. E-mail: nandk@kaist.ac.kr; Tel: +82-42-350-1156*
[b] *Department of Chemistry, University of Warsaw, Warsaw, 02-089, Poland. E-mail: gorecka@chem.uw.edu.pl; Tel: +48-22-8220211*
[c] *Pohang Accelerator Laboratory, POSTECH, Pohang, 790-784, Republic of Korea*
[d] *UNIST Central Research Facilities, UNIST, Ulsan 689-798, Republic of Korea*
[e] *Department of Chemistry and Biochemistry, University of Maryland, College Park, MD 20742, USA*

Dong Ki Yoon (nandk@kaist.ac.kr)

Graduate School of Nanoscience and Technology (WCU),

KAIST, Daejeon 305-701

Republic of Korea.

Tel: 82 42 350 1116 FAX: +82 42 350 1110

http://yoon.kaist.ac.kr

# Abstract

We studied the correlation between the molecular structure and the formation of helical nanofilaments (HNFs) of bent-core dimeric molecules with varying linkage lengths. To obtain precise structural data, a single domain of HNFs was prepared under physical confinement using porous 1D nanochannels, made up of anodic aluminium oxide films. Electron microscopy and grazing incidence X-ray diffraction were used to elucidate the linkage length-dependent formation of HNFs.

Chirality is one of the most interesting topics in fundamental sciences as well as in application-related research. Chiral liquid crystalline (LC) materials, in particular, have been used in a broad range of applications such as chiral sensors and charge transporting media owing to their strong and fast response to various external stimuli.[1–5] An example of the chiral LC phases is the B4 phase (also referred to as the helical nanofilament- HNF phase), built of twisted layers, and xhibiting macroscopic chirality and polarity that are useful for chiral sensors,[5] nanofabrication,[6,7] or non-linear optical applications.[8,9] The structure of the B4 phase has been discussed in many previous studies;[10–14] it is a lamellar crystal that forms helical nanofilaments and such an unusual morphology is believed to be related to symmetry-breaking in the packing of twisted molecular conformers that induces the saddle-splay deformation of molecular layers. One of the basicmechanisms responsible for the formation of the lamellar structures, e.g. smectic LC phases, is the nanosegregation of different molecular units, as well as rigid aromatic and flexible aliphatic parts, into sub-layers. Each part contributes to the molecular organisation during the formation of the layered structure, which mostly depends on a balanced interaction between the rigid core units and the flexible alkyl tail or spacer chains.[15–19] In determining the correlation between the molecular structure and macroscopic functionalities of HNFs (i.e. chirality and polarity), considerable efforts have been undertaken to elucidate the roleof the particularmolecular components (especially the alkyl chain attached to the molecule).[20–22] In spite of these efforts, the layering mechanism of the HNF phase is still not clearly understood, as a result of the inherent structural complexity. Most of the materials exhibiting a B4 phase are rigid bent-core molecules, which obtain their bent molecular geometry via a 1,3 substitution of the central phenyl ring. In the present study, we focus on another class of bent molecules, i.e. flexible dimers, for which the formation of the B4 phase is rather rare.[23,24] Four homologues with different numbers of carbon atoms in their respective spacers, located between the two rigid linear cores, were studied (Fig. 1a).[23] For the precise structural analysis of the molecular organisation of such a system, the LC materials were introduced into a nanoconfined geometry using porous anodic aluminium oxide (AAO) film,[25] which is the most effective tool in controlling the growth of a single HNF. Scanning electron microscopy (SEM) was used to analyse the morphological changes of the nanoconfinedHNFs, and details of themolecular organization within the layers were investigated using the grazing incidence X-ray diffraction (GIXD) technique.

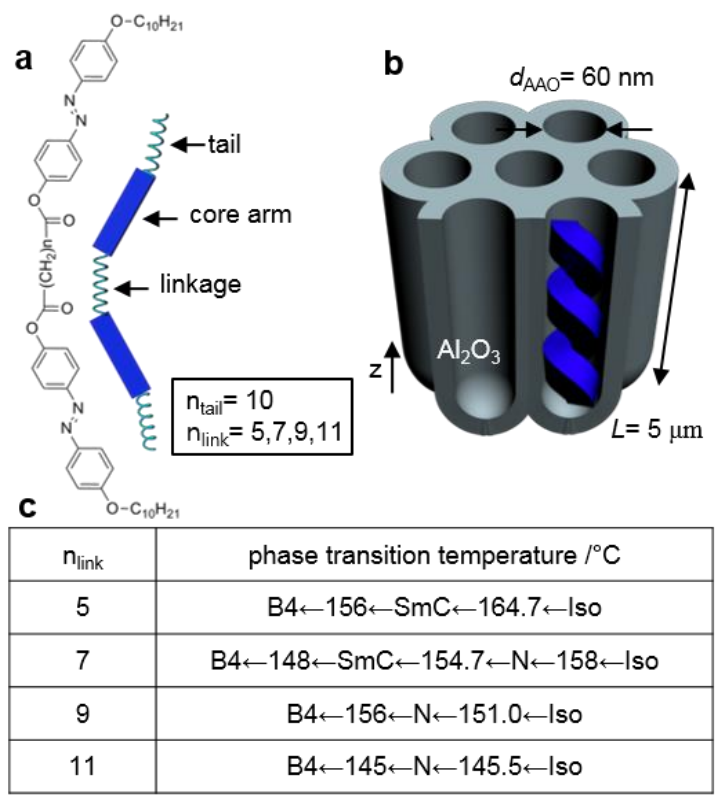

**Fig. 1** Materials and experimental conditions (a) Molecular structure and the representative description of bent-shaped dimeric LC molecules, with varying lengths of linking spacer ($n_{link}$ = 5-11) with the fixed tail ($n_{tail}$ = 10). (b) Geometry of the 1D porous oxide nanochannels (AAO) having a fixed dimension of $d_{AAO}$ = 60 nm and $L$ = 5 μm. (c) Thermal phase transition information.

## Results and discussion

All homologues have the same rigid cores and terminal alkyl tails ($-C_{10}H_{21}$), differing only in the internal spacer chain length, $-(CH_2)n-$ from $n_{link}$ = 5 to 11 (Fig. 1a). Only molecules having an odd number of spacer carbon atoms were studied as the even-numbered homologues did not reveal the B4 phase; instead they formed smectic C and J phases similarly to typical rod-like molecules.[23] Apart from the B4 phase short homologues, with $n_{link}$ = 5 and 7, exhibited the smectic C phase at a higher temperature range, while for longer ones the nematic phase was observed (Fig. S1, ESI†).

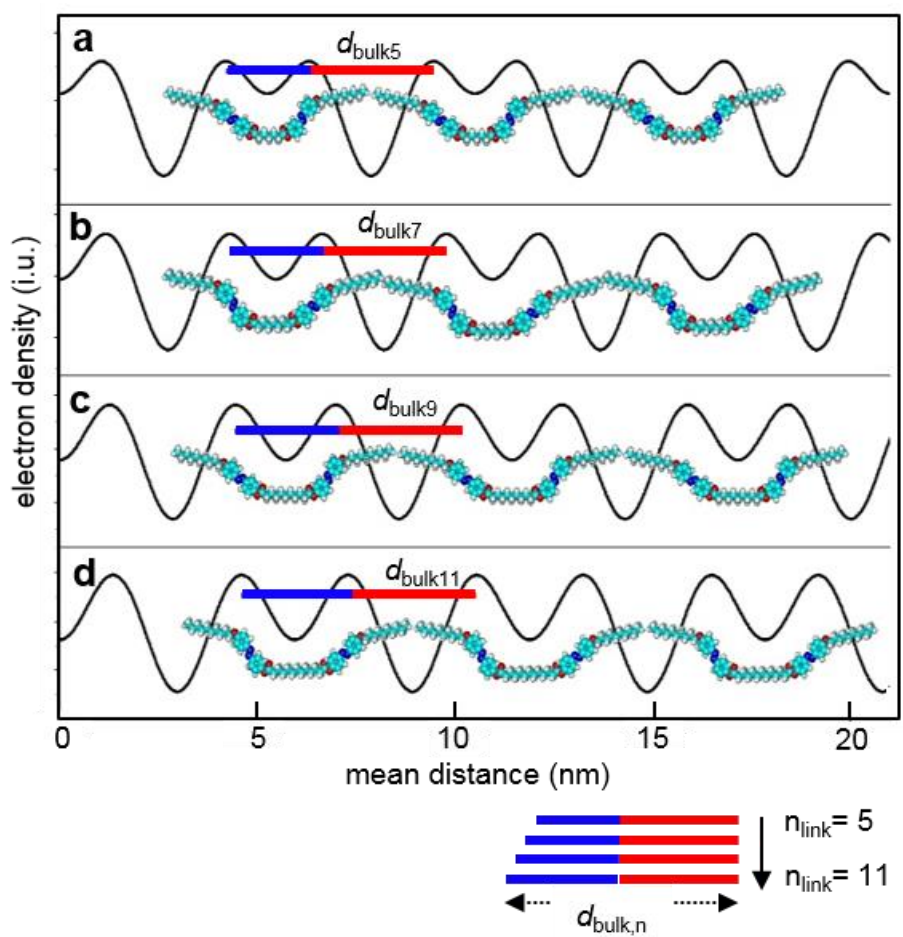

**Fig. 2** Electron density profiles along the layer normal determined from the powder X-ray diffraction data by the reversed Fourier transform. The measurements were performed for the bulk samples at room temperature. With elongation the intrmolecular spacer the dip in the electron density in the middle of the layer becomes more pronounced and wider, reflecting increasing distance between two mesogenic cores within single molecule (blue line). The distance between mesogenic cores fron neighbouring layes (red lines) is the same for all homologues.

In order to see themolecular order, X-ray diffraction experiments were performed for the bulk samples at room temperature, revealing the lamellar crystalline structure with d-spacing that

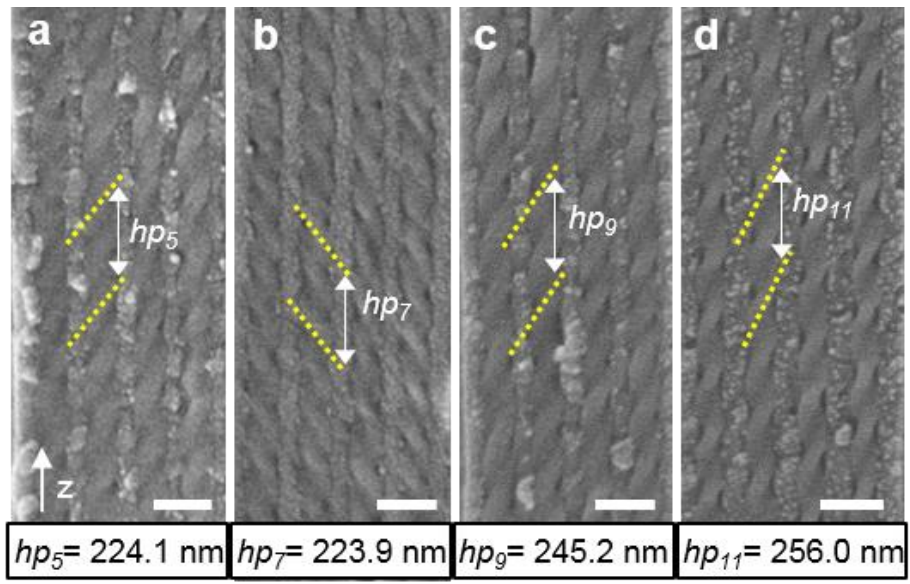

**Fig. 3** Morphologies of nanoconfined HNFs for homologues having different linkages. (a) $n_{link}$ = 5, (b) $n_{link}$ = 7, (c) $n_{link}$ = 9, and (d) $n_{link}$ = 11, respectively. Scale bars: 100 nm.

corresponds to a molecular length of $d_{nL}$ (Fig. 2). Electron density profiles determined by reversed Fourier transform from the diffraction peak intensities showed that an additional electron density minimum exists in the middle of the layer, corresponding to the position of the internal alkyl spacer of the molecules, and that the minimum was better defined for longer spacer homologues. The morphology of bulk samples was studied using AFM, in which the bulk B4 phase showed strongly distorted lamellar crystallites with clearly visible regions of saddle splay deformations (Fig. S1, ESI†), with the formation of randomly grown rope-like filaments, having a width ($w$) of ~50 nm and a helical pitch ($hp$) of ~150 nm. When the materials are placed under spatially limited conditions, i.e. in 1D-nanochannels,[25] the molecules self-restricted into uniformly twisted layers because of the combination of heat release at the phase transition and the reservoir-like big isothermal boundary condition. To obtain the vertically aligned individual HNFs, all LC samples in the isotropic melt state (175 °C) were loaded into the AAO nanochannels that have a fixed diameter of $d_{AAO}$ = 60 nm and a length of ~5 μm (Fig. 1b). The samples were then slowly cooled to room temperature at a steady cooling rate of 1 °C

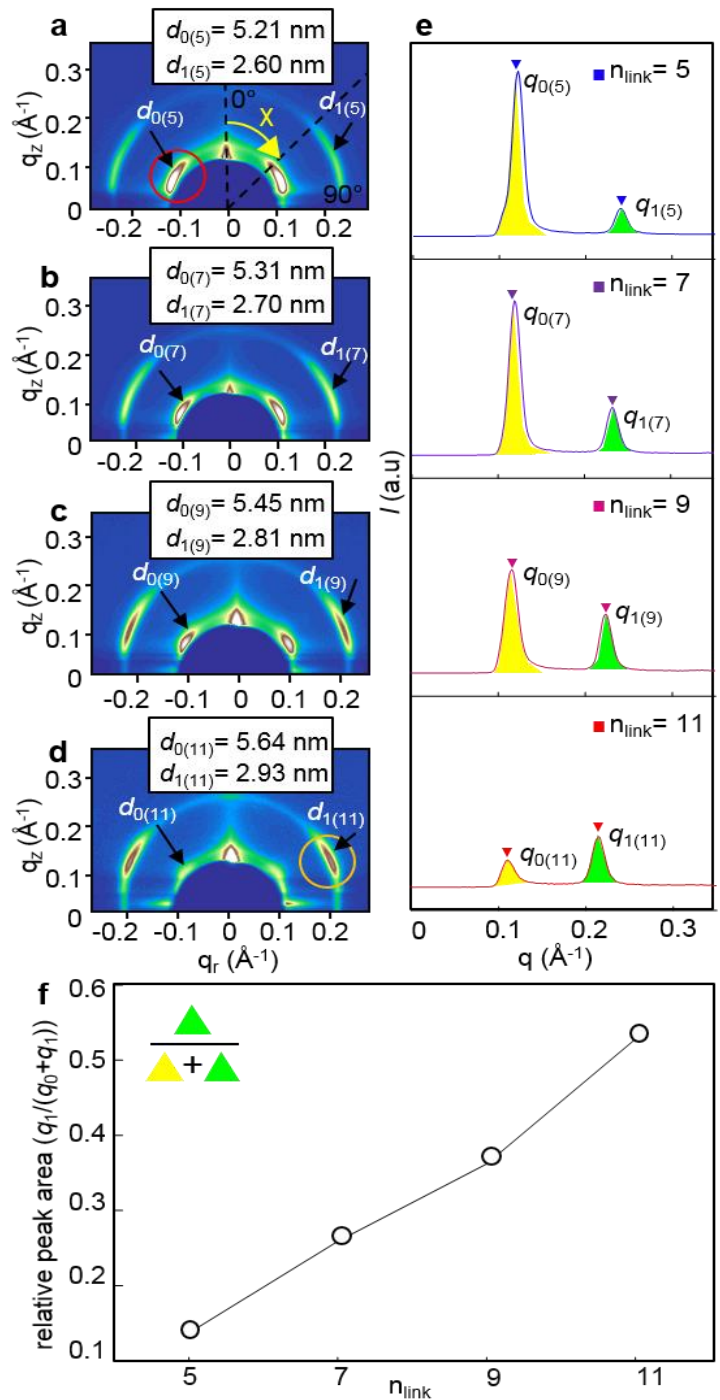

**Fig. 4** GIXD patterns of homologues in confined geometry ($d_{AAO}$= 60 nm). (a-d) 2D-GIXD patterns acquired in the small angle region ($q$ = 0- 0.3 Å$^{-1}$), and their 1D profiles (e). (f) Relative peak area integrated from (e) the ratio of the first order peak $q_{0(n)}$ to the second order peak $q_{1(n)}$, which shows a linear relationship with increasing the linkage length.

/min. The HNF structure of each homologue in AAO was directly visualised by SEM (Fig. 3). All homologues self-stabilised into similar morphologies at room temperature regardless of their linkage length, with the vertical HNFs exhibiting clear macroscopic chirality and uniformity. Apparently, a strong thermal gradient controlled the growth direction of HNFs in

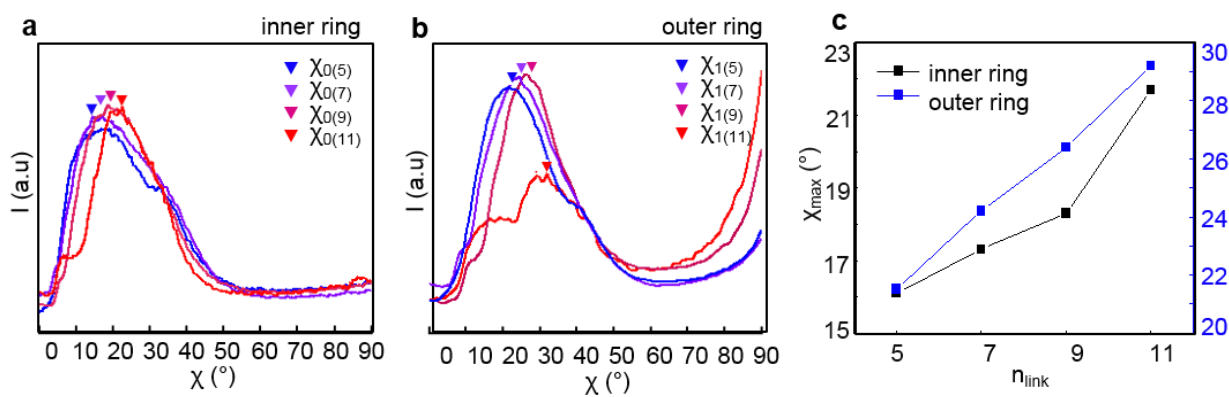

Fig. 5 Azimuthal angle scan analysis at the small-angle region. The maximum angle deviation ($\chi_{max}$) from the filament axis $f$ (off-center, 0°) increases gradually from $\chi_{0(5)}$ = 16.1° to $\chi_{0(5)}$ = 21.7° in the $q$-range of layer packing region (a), and from $\chi_{1(5)}$ = 21.5° to $\chi_{1(11)}$ = 29.2° in the second order peak region (b) as the alkyl spacer length increases from $n_{link}$ = 5-11. (c) In both regions, a linear change of $\chi_{max}$ over an increasing linkage length is shown.

the AAO nanochannels; the HNFs were nucleated along the vertical direction, from the top to the bottom of the AAO film, as heat was applied from underneath the channels. The HNFs were individually grown with fixed handedness in a single pore, but the handedness of neighbouring HNFs was randomly distributed because of the isolated and confined conditions. When the structure of the HNFs was examined more closely, it was discovered that the helical pitch, $hp$, is also dependent on the molecular length. The $hp$ value increases proportionally to the increase of the length of the spacer chain, which implies that the elastic cost of layer twisting must be related to the molecular structure, i.e. it is easier to twist layers made up of shorter spacer molecules. This will be further discussed in the later section with Fig. 5. In order to analyse the layer ordering and orientation of the samples in a confined geometry, the GIXD experiments were carried out with synchrotron radiation (Fig. 4, 5 and Fig. S2, ESI†). Fig. 4 shows information associated with the HNF layer orientation in the small-angle region ($q <$ 0.35 Å$^{-1}$). The azimuthal positions of the diffraction patterns showed a diagonal layer arrangement of HNFs as reported by previous studies (Fig. 4a–d).[12,25] The circular-averaged 1D profiles (Fig. 4e) show two peaks in different q ranges, around 0.12 Å$^{-1}$ and 0.23-0.24 Å$^{-1}$, respectively, for all homologues. These peaks do not have strictly identical azimuthal positions (Fig. 5). Based on the peak positions at presented $q$ values, the first order peaks for the layer thickness $d_{0(n)}$ appeared in the inner circle region (Fig. 4), corresponding to the bulk state observation (Fig. 2), whereas the 2nd order harmonic $d_{1(n)}$ was observed in the outer circle. As the linkage length was elongated from $n_{link}$ = 5 to n = 11, the value of $d_{0(n)}$ increased from 5.13 to 5.64 nm, which is reasonable considering that every two additional carbon atoms result in about ~0.23 nm increase in length. The intensity and integrated area of diffraction patterns corresponding to $d_{0(n)}$ (the first order) and $d_{1(n)}$ (the second order) are dramatically influenced by an increasing $n_{link}$. This trend is summarised in Fig. 4f, which shows a linear increment in the ratio of integrated areas for the diffraction intensities of $d_{1(n)}$ to $d_{0(n)}$. To explain this phenomenon, we consider the basic information of the electron density distribution in layers with the bulk reference in Fig. 2. With elongation of the internal alkyl linkage in dimers, the electron density along the layer normal is changed as the inter-core distance changes, in which the additional density minimum becomes deeper and its width increases (core-linkage-core distance, blue lines in Fig. 2), whereas the inter-molecular distance between electron density maxima measured through the inter-layer interface is the same for all homologues (core-tail–tail-core distance, red lines in Fig. 2). At the point where the spacer is comparable to that of the tail length ($n_{link}$ = 11), the periodicity of electron density along the layer normal is approaching $d/2$, leading to a diminishing intensity of the first order diffraction peak. The different molecular organisations observed in the layered structures of each homologue are also related to the change in rigidity and elastic properties of crystal layers in HNFs, which can be determined by the changes in the twisting angle and pitch of HNFs (with these parameters, in turn, being dependent on the linkages) (Fig. 3).[26–30] Fig. 5 shows the 1D-angular scan for the diffraction peak intensities in the small angle region, illustrating the twist angle variation in the layers in HNFs, which is denoted as $w$ (i.e., based on deviation from the helical axis z). As shown in Fig. 5a and b, the w value at each maximum peak intensity is increased from $\chi_{0(5)}$ = 16.1° to $\chi_{0(11)}$ = 21.7° in the HNF layers (the first order-inner ring), and from $\chi_{1(5)}$ = 21.5° to

$\chi_{1(11)}$ = 29.2° in the second order peak (outer ring). In both regions, the value of w shows proportional changes with the elongation of linkage chains, as summarised in Fig. 5c. As all the homologues were under the same spatial conditions with a fixed dimension AAO wall ($d_{AAO}$ = 60 nm), it is assumed that the surface anchoring force from the channel was applied equally to the LC material, and the same number of layers existed in each single pore (N~10 of 5–6 nm thick layers, with a 2D-crystalline order). These results agree with the morphological changes determined from the *hp* values (Fig. 3). The hp distance is linearly increased with the elongation of the spacers, suggesting that the layers were less twisted with a longer spacer chain. To explain this, we consider the molecular configuration change over the phase transitions: in layers of tilted bent-core molecules, top and bottom arms in sub-layer tilt almost orthogonal to each other while each sub-layer tends to dilate along its tilt direction.[12] Such an intra-layer structural mismatch drives the formation of HNFs and determines the helical pitch of HNFs. When the length of the spacer linker increases, it is reasonable that the mismatch is released to some extent and it favours a smaller curvature and a larger pitch. Therefore, such different layering behaviour is assumed to be responsible for the change in the layer twisting angle, which leads to the different value of helical pitch.

Conclusions

We investigated how the linkage variation in the dimer type bent-core molecules gives rise to different layering behaviours during the formation of helical nanostructures. Under nanoscale spatially confined conditions, single domains of HNFs were prepared to analyse their structural changes using SEM and GIXD. It was found that the elongation of the alkyl linkage spacer between two rigid molecular cores influences the elastic properties of HNFs, i.e. the twist angle and helical pitch of the chiral nanofilaments. This result facilitates current understanding of the underlying mechanism involved in the formation of helicallayered structures, resulting from the specific arrangements of aromatic cores and aliphatic spacers in molecules.

# Experimental

A series of bent-shaped LC homologues (4-decyloxy-40-hydroxybiphenyl with a flexible alkyl linkage) was prepared according to a method previously reported in the literature.[23,24] The bulk morphologies of each material were observed in their bulk state, and they were prepared within two PI pre-coated glass substrates that were rubbed along a single direction. LC materials were loaded into the cells at 170 1C (at which the molecules are at an isotropic melt state) by capillary force. Sequential phase behaviours were observed with polarized optical microscopy (POM), while undergoing a gradual cooling process at a fixed cooling rate of 5 °C/min. Nanoconfined HNFs of these chemical homologues were acquired using a simple nanoconfining process utilising AAO. Cylindrical AAO nanochannels were prepared by a conventional two-step anodisationmethod.[31] A high-purity annealed aluminium foil (99.99%, Alfa Aesar) was electro-polished in amixed solution of perchloric acid and ethanol, and then anodised in a 0.3 M oxalic acid at a voltage of 40 V for 7 h (at 10 °C). Pre-grown initial oxide layers were chemically etched using a mixture of chromic acid and phosphoric acid, before undergoing a second anodisation process under the same experimental conditions for 1 h. Regularly grown oxide pores were widened in a phosphoric acid solution at 38 °C for 20 min, leading to the formation of highly uniform AAO nanochannels ($d_{AAO}$ = 60 nm and $L$ = 5 mm). With the prepared 1D nanochannels, the LC material (at the isotropic melt state) was injected into porous AAO at 170 °C by capillary force and cooled to room temperature at a cooling rate of 5 °C min. After finishing the whole phase transitions, the residual bulk HNFs were removed by physical etching, giving only the nanoconfined HNFs for SEM and GIXD analysis. Details of the GIXD setup are described in the ESI.†

# Acknowledgements

This work was financially supported by the National Research Foundation (NRF) of the Korean Government (MOE: 2014- S1A2A2027911). AZ, DP and EG were financially supported by NCN (Poland) program no. UMO-2012/07/B/ST5/02448. The experiments at the PLS-II were supported in part by MSIP and POSTECH.

# Notes and references

# Supplementary Materials for

# Linkage-length dependent layering behavior of bent-core molecules in helical nanostructures

Hanim Kim,[a] Anna Zep,[b] Seong Ho Ryu,[a] Hyungju Ahn,[c] Tae Joo Shin,[d] Sang Bok Lee,[e] Damian Pociecha,[b] Ewa Gorecka*[b] and Dong Ki Yoon*[a]

[a] *Graduate School of Nanoscience and Technology, KAIST, Daejeon 305-701, Republic of Korea. E-mail: nandk@kaist.ac.kr; Tel: +82-42-350-1156*
[b] *Department of Chemistry, University of Warsaw, Warsaw, 02-089, Poland. E-mail: gorecka@chem.uw.edu.pl; Tel: +48-22-8220211*
[c] *Pohang Accelerator Laboratory, POSTECH, Pohang, 790-784, Republic of Korea*
[d] *UNIST Central Research Facilities, UNIST, Ulsan 689-798, Republic of Korea*
[e] *Department of Chemistry and Biochemistry, University of Maryland, College Park, MD 20742, USA*

*This material contains:*

*Figures S1-S2*

*Figure S1*

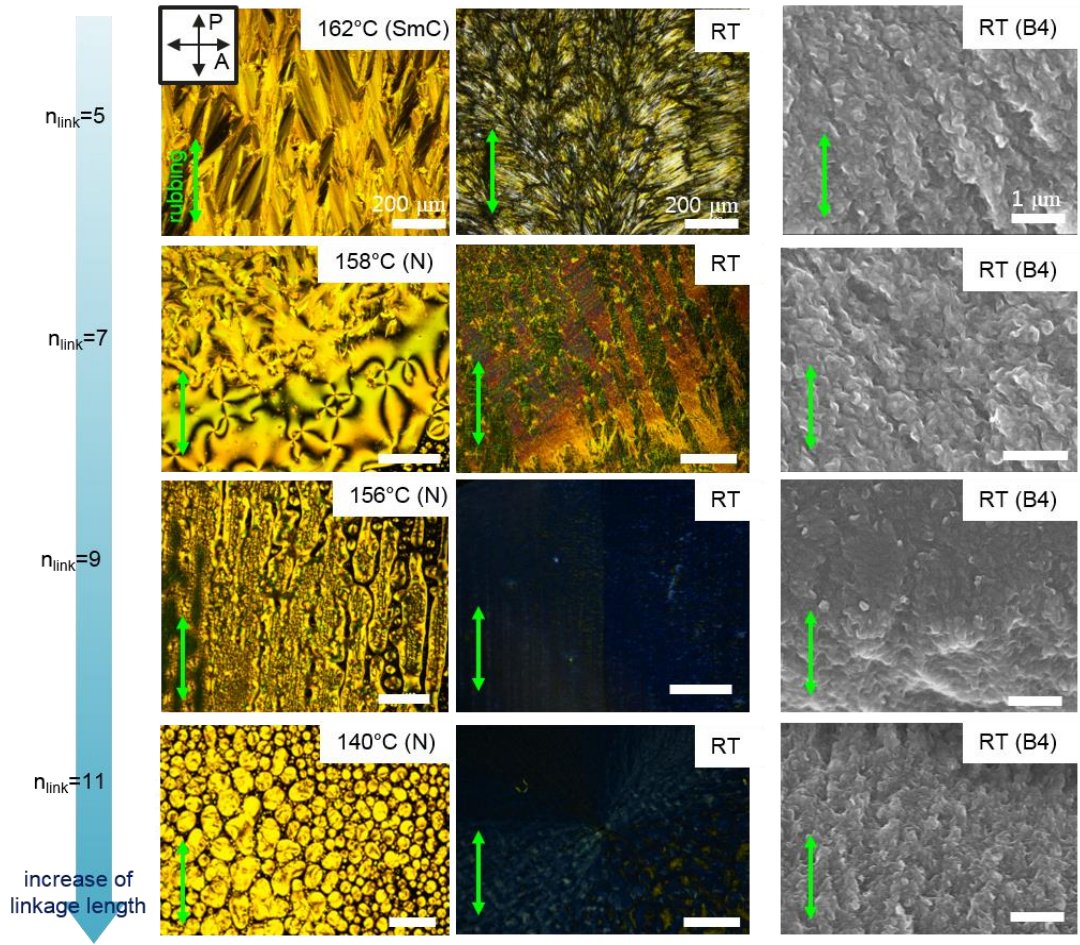

S1. Observations by DRLM and SEM on the overall thermal transitions of bulk HNFs with the homologues upon cooling (5 °C/min), which imply how the filaments are grown from preceding LC phases (Smetic C or Nematic) before reaching to B4, where each homologue alters molecular organization in layering. For the case of short spacer ($n_{link}$= 5 and 7), it shows the pre-smectic feature at high temperature range while the long molecules ($n_{link}$= 9 and 11) only reveal nematic phase in the whole temperature range before B4, meaning that the long-linkage group might induce geometric overlapping among the molecules which is evident in the polarized optical textures. All cases ternimate their thermal transitions for B4 phase at room temperature, which show random orientation in morphology regardless of the pre-rubbed direction (green arrow). All scale bars, 200μm.

*Figure S2*

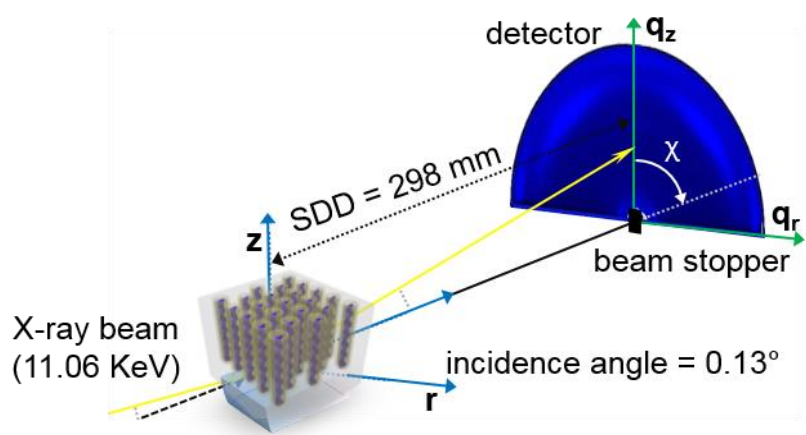

S3. Geometry of grazing incidence X-ray diffraction measurement for the simultaneous investigation of the molecular orientation and layer arrangement. X-ray beam is incidentally guided to the sample, having the beam energy of 11.06 KeV, and the size of 700 μm (vertical) by 300 μm (horizontal). SDD (Sample to detector distance) was 298 mm with a two-dimensional charge-coupled device (2D-CCD) detector.